\documentclass[a4paper,11pt]{article}
\pdfoutput=1
\usepackage{jcappub}
\usepackage[T1]{fontenc}
\usepackage{graphicx}
\usepackage{xcolor}
\usepackage[normalem]{ulem}
\usepackage{booktabs}
\usepackage{array}
\usepackage{tikz}

\newcommand{\plin}{P_{\rm lin}}
\newcommand{\kk}{\mathbf{k}}
\newcommand{\qq}{\mathbf{q}}
\newcommand{\pp}{\mathbf{p}}
\newcommand{\dd}{\mathrm{d}}
\newcommand{\re}{\operatorname{Re}}
\newcommand{\im}{\operatorname{Im}}
\newcommand{\intq}{\int_{\qq}}
\newcommand{\intp}{\int_{\pp}}
\newcommand{\hmpc}{\,h\,{\rm Mpc}^{-1}}

\title{Two-loop matter power spectrum from FFTLog and kite integrals}

\author[a,b,1]{Paola Cappellazzi,\note{Corresponding author.}}
\author[a,b]{Johannes M. Henn,}
\author[c,d]{Marko Simonovi\'c}

\affiliation[a]{Max-Planck-Institut f\"ur Physik (Werner-Heisenberg-Institut),\\ Boltzmannstra\ss e 8, 85748 Garching, Germany}
\affiliation[b]{Max Planck--IAS--NTU Center for Particle Physics, Cosmology and Geometry}
\affiliation[c]{Dipartimento di Fisica e Astronomia, Universit\`a degli Studi di Firenze,\\ Via G.\ Sansone 1, I-50019 Sesto Fiorentino, Italy}
\affiliation[d]{INFN, Sezione di Firenze,\\ Via G.\ Sansone 1, I-50019 Sesto Fiorentino, Italy}

\emailAdd{pcappe@mpp.mpg.de}
\emailAdd{henn@mpp.mpg.de}
\emailAdd{marko.simonovic@unifi.it}

\abstract{
{We present a fast and accurate method for computing the two-loop matter power spectrum in the Effective Field Theory of Large-Scale Structure (EFTofLSS). Using FFTLog, we decompose the linear power spectrum into complex power laws, reducing the two-loop corrections to cosmology-independent massless Feynman integrals of kite topology. A central numerical challenge is the evaluation of these integrals for complex powers with large imaginary parts, where the highly oscillatory integrands make accurate numerical evaluation particularly challenging. We overcome this difficulty using a Mellin–Barnes representation summed over a uniform lattice, achieving approximately ten digit precision even for large imaginary powers, with an error estimate for each integral. The resulting precomputed matrices allow the two-loop power spectrum to be evaluated for a new cosmology in a few milliseconds. Comparing with direct numerical integration, we find agreement at the $10^{-4}$ level relative to the power spectrum up to $k=0.2\hmpc$ after accounting for EFTofLSS counterterm contributions and FFTLog decomposition errors. The same framework extends to biased tracers and redshift-space distortions, providing the building blocks for general two-loop calculations in EFTofLSS. We make the code for evaluating the kite integrals publicly available, together with precomputed matrices and an implementation for the two-loop matter power spectrum as a concrete application. Our results can be directly incorporated into existing FFTLog-based codes for large-scale structure, extending their current one-loop capabilities to two loops within the same computational framework.}

}


\makeatletter\g@addto@macro\beforetochook{\enlargethispage{2\baselineskip}}\makeatother
\begin{document}
\maketitle
\flushbottom

\section{Introduction}
\label{sec:intro}

Galaxy surveys such as DESI and Euclid~\cite{DESI:2016fyo,Amendola:2016saw} measure the clustering of matter with percent-level precision on scales where the density fluctuations are still small, but gravity already couples different scales. Full-shape analyses of these data use the effective field theory of large-scale structure (EFTofLSS)~\cite{Baumann:2010tm,Carrasco:2012cv}, which builds on standard perturbation theory~\cite{Bernardeau:2001qr} and adds counterterms for the effect of short-scale physics. 
{One-loop analyses are now standard~\cite{Ivanov:2019pdj,DAmico:2019fhj,DESI:2024jxi} and rely on fast codes for the loop integrals, such as FAST-PT~\cite{McEwen:2016fjn,Fang:2016wcf}, CLASS-PT~\cite{Chudaykin:2020aoj}, PyBird~\cite{DAmico:2020kxu}, velocileptors~\cite{Chen:2020fxs}, FOLPS~\cite{Noriega:2022nhf}, CLASS-OneLoop~\cite{Linde:2024uzr}, and neckar~\cite{Harder:2026uei}; see also refs.~\cite{delaBella:2017qjy,Maus:2024sbb}.}
Two-loop corrections extend the theory to smaller scales and quantify the theory error of one-loop analyses~\cite{Carrasco:2013sva,Carrasco:2013mua,Baldauf:2015aha,Foreman:2015lca}.

Each loop adds a three-dimensional integral over an unobserved wavenumber, and the integrand contains several factors of the linear power spectrum $\plin$, which for a realistic cosmology is known only numerically~\cite{Blas:2011rf}. In a cosmological analysis these integrals have to be repeated at every point in parameter space~\cite{Cataneo:2016suz}. The most direct way integrates over the loop momenta numerically, with the integrand arranged so that large infrared contributions cancel point by point~\cite{Carrasco:2013sva,Blas:2013aba}; the two-loop power spectrum has been computed in this way with full time dependence, in redshift space and with local subtraction of singularities~\cite{Garny:2020ilv,Garny:2022fsh,Taule:2023izt,Anastasiou:2025jsy}, and the three-loop spectrum has been obtained~\cite{Blas:2013aba,Konstandin:2019bay}. Fast methods instead expand $\plin$ in a fixed set of functions, so that each loop correction becomes a sum of cosmology-independent integrals, computed once, with coefficients that carry all the dependence on cosmology: complex power laws (FFTLog)~\cite{Hamilton:1999uv,McEwen:2016fjn,Schmittfull:2016jsw,Schmittfull:2016yqx,Fang:2016wcf,Simonovic:2017mhp}, massive propagators~\cite{Anastasiou:2022udy}, or a basis optimized for a range of cosmologies~\cite{Bakx:2024zgu,Bakx:2025pop}, recently applied to the two-loop power spectrum~\cite{Bakx:2025jwa}. Bases fitted to the spectra of interest need few functions and therefore few integrals. Complex power laws need more, typically several tens, but they are not tied to any cosmology: the same precomputed integrals apply to any linear spectrum that the power laws resolve, and they also evaluate the integrals of an optimized basis~\cite{Bakx:2025pop}.

In the FFTLog approach to loop integrals of large-scale structure~\cite{Simonovic:2017mhp}, each power law acts as a massless propagator raised to a complex power, so every loop integral becomes a cosmology-independent massless Feynman integral in three dimensions. At one loop these integrals are ratios of Gamma functions, and the method is in production use~\cite{Chudaykin:2020aoj}. At two loops they reduce to the massless kite integral with complex powers; this reduction, series representations of the kite integral and the full set of two-loop formulas were given in ref.~\cite{Simonovic:2017mhp}. The remaining difficulty is the evaluation of the kite integrals at the powers that FFTLog requires. Their imaginary parts are fixed by the frequency grid and reach values of order ten or more (section~\ref{sec:fftlog}); at such values the integrands oscillate rapidly, and the region where they are large moves away from the origin. The M.Sc. thesis of one of us~\cite{Cappellazzi:2026msc} showed that most of the kite integrals have closed forms and derived, for the others, a two-fold Mellin--Barnes (MB) representation at complex powers.

Here we turn these results into precomputed matrices for the complete two-loop matter power spectrum and test them. We evaluate the MB representation on straight contours by summing the integrand over a lattice of equally spaced points, and continue it analytically to all powers~\cite{Czakon:2005rk}. For an integrand that is analytic near the contour and decays fast, this sum converges exponentially~\cite{Trefethen:2014}. Large imaginary parts of the powers only move the region where the integrand is large, without slowing its decay, so the lattice captures that region automatically, whereas an adaptive integrator can miss it (section~\ref{sec:numerics}). We compute the matrices for a grid that is able to resolve a realistic linear spectrum with baryon acoustic oscillations (BAO), the finest feature of the spectrum.

Our main result is the two-loop power spectrum computed from these matrices, compared with a direct numerical integration of the physical loop integrals (section~\ref{sec:spectrum}). The comparison has to account for what FFTLog computes. The sum of power laws represents the linear spectrum on a finite interval of wavenumbers and continues it periodically outside, and the loop integrals are defined by analytic continuation in the powers, which sets power-law divergences at large loop momenta to zero~\cite{Simonovic:2017mhp,Pajer:2013jj}. FFTLog thus evaluates the loop integrals in analytic regularization. Loop momenta far below or far above $k$, which are the only ones affected, contribute in universal forms~\cite{Goroff:1986ep,Jain:1995kx,Scoccimarro:1995if,Carrasco:2013mua,Baldauf:2015aha}: the infrared contributions cancel in the sum of all two-loop terms, and the ultraviolet ones have the form of the counterterms of the EFTofLSS, with coefficients that we compute from the moments of the spectrum outside the interval and from the large-momentum limits of the integrands. This is the familiar statement that the loop integrals depend on wavenumbers where perturbation theory fails and that the counterterms absorb this dependence~\cite{Baumann:2010tm,Carrasco:2012cv,Anastasiou:2025jsy}. After these terms are accounted for, the FFTLog spectrum agrees with the direct integration to {better than $10^{-4}$} of the linear spectrum up to $0.2\hmpc$. The remaining difference, there and at larger $k$, is that between the input spectrum and its representation by power laws. Different biases give the same result at this level. For one contribution, $P_{33\text{-}\rm II}$, we give a detailed example of how these issues are treated (appendix~\ref{sec:allchecks}).

Section~\ref{sec:kite} treats the kite integral independently of cosmology. Section~\ref{sec:pk} sets up the power spectrum, its precomputed matrices and the analytic regularization. Section~\ref{sec:spectrum} presents the two-loop spectrum, the direct integration and the comparison, and the accuracy and the cost of the method; section~\ref{sec:discussion} gives applications and conclusions. Appendix~\ref{app:hyp} gives the check of our program against hypergeometric representations, appendix~\ref{app:lit} the counterterm-form terms explicitly and the comparison of our result with published two-loop results, and appendix~\ref{sec:allchecks} the example of $P_{33\text{-}\rm II}$.

\section{The kite integral at complex powers}
\label{sec:kite}

\subsection{Definition and known results}
\label{sec:def}
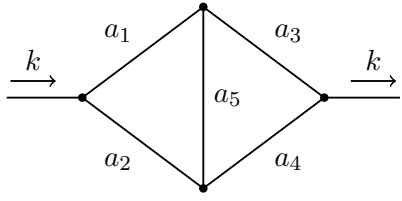
\begin{figure}[t]
\centering
\begin{tikzpicture}[line width=0.7pt,baseline=(L)]
\coordinate (L) at (0,0);
\coordinate (R) at (3.2,0);
\coordinate (T) at (1.6,1.2);
\coordinate (B) at (1.6,-1.2);
\draw (-1.0,0) -- (L);
\draw (R) -- (4.2,0);
\draw[->] (-0.95,0.22) -- (-0.35,0.22) node[midway,above] {$k$};
\draw[->] (3.55,0.22) -- (4.15,0.22) node[midway,above] {$k$};
\draw (L) -- (T) node[midway,above left] {$a_1$};
\draw (L) -- (B) node[midway,below left] {$a_2$};
\draw (T) -- (R) node[midway,above right] {$a_3$};
\draw (B) -- (R) node[midway,below right] {$a_4$};
\draw (T) -- (B) node[midway,right] {$a_5$};
\foreach \v in {L,R,T,B} \fill (\v) circle (1.6pt);
\end{tikzpicture}
\caption{The kite integral of eq.~\eqref{eq:kite}. Line $i$ carries the power $a_i$; with the external momentum $k$ entering on the left, lines 1 to 5 carry the momenta $q$, $k-q$, $p$, $k-p$ and $q-p$. Lines 1, 2 and 5 carry the loop momentum $q$; they join the left external vertex and the two internal vertices and form the triangle of section~\ref{sec:MB}.}
\label{fig:kite}
\end{figure}

The kite integral is the two-loop self-energy integral with five massless propagators: two lines leave each external vertex, and a fifth line joins the two internal vertices. With an arbitrary complex power $a_i$ on each propagator, we normalize it in $d$ Euclidean dimensions as
\begin{equation}
I(a_1,\ldots,a_5;d)=\int\frac{\dd^dq\,\dd^dp\;|\kk|^{2(\sum_ia_i-d)}\,\pi^{-d}}{|\qq|^{2a_1}\,|\kk-\qq|^{2a_2}\,|\pp|^{2a_3}\,|\kk-\pp|^{2a_4}\,|\qq-\pp|^{2a_5}}\,,
\label{eq:kite}
\end{equation}
with $k=|\kk|$, so that it depends only on the powers $a_i$ and the dimension $d$. Figure~\ref{fig:kite} shows the labeling of the lines. For example, the integral with unit powers in four dimensions is $I(1,1,1,1,1;4)=6\zeta_3$~\cite{Chetyrkin:1980pr}. In the context of the EFTofLSS, $d=3$.

The integral converges when the real parts of the powers satisfy the power-counting conditions
\begin{equation}
\begin{gathered}
\re a_i<\tfrac d2\;(i=1,\ldots,5)\,,\qquad \re a_{135}<d\,,\qquad \re a_{245}<d\,,\\
\re a_{125}>\tfrac d2\,,\qquad \re a_{345}>\tfrac d2\,,\qquad \re a_{1234}>\tfrac d2\,,\qquad \re a_{12345}>d\,,
\end{gathered}
\label{eq:conv}
\end{equation}
with $a_{i\cdots j}\equiv a_i+\dots+a_j$; the first line comes from small and the second from large loop momenta. Elsewhere it is defined by analytic continuation. The kite integral is invariant, up to Gamma-function factors, under a group of 1440 transformations of the powers~\cite{Broadhurst:1986bx,Barfoot:1987kg}. Among these are the 24 permutations of the vertices of the tetrahedron obtained by closing the kite with a sixth line between the external vertices, with power $a_6=\tfrac{3d}2-a_{12345}$, and the same permutations combined with the reflection $a_i\to\tfrac d2-a_i$. They relate a kite integral to others that may be easier to evaluate, and they serve as checks (section~\ref{sec:checks}).

The two-loop kite integral has been studied extensively in the literature~\cite{Grozin:2012xi,Kotikov:2018wxe}. For integer powers, integration-by-parts identities~\cite{Chetyrkin:1981qh,Tkachov:1981wb} reduce it to products of one-loop bubbles. For non-integer powers, closed-form expressions exist only for special families. In particular, single-series representations have been derived for $I(1,1,1,1,a_5;d)$~\cite{Kazakov:1983pk}, and for configurations in which the two powers associated with edges meeting at an internal vertex both equal $\tfrac d2-1$~\cite{Kotikov:1995cw}. For generic powers it has been written as a double series~\cite{Bierenbaum:2003ud,Simonovic:2017mhp,Derkachev:2020xib}. Such series representations however are impractical for our purposes, as they often converge too slowly, or not at all. We therefore use them only as cross-checks (appendix~\ref{app:hyp}), and instead evaluate the kite integral numerically from its Mellin--Barnes representation.

\subsection{Mellin--Barnes representation}
\label{sec:MB}
MB representations recast Feynman integrals as contour integrals involving ratios of Gamma functions~\cite{Smirnov:1999gc,Tausk:1999vh,Dubovyk:2022obc}. For the evaluation of the kite integral, we adopt the MB representation that was independently derived in the M.Sc. thesis~\cite{Cappellazzi:2026msc}; Bierenbaum and Weinzierl~\cite{Bierenbaum:2003ud} gave an equivalent formula. It follows by performing the loop integrations one at a time. First, the 125-triangle is written in Feynman parameters, and its denominator is factorized by introducing a two-fold MB integral. The remaining integrals, including the bubble of lines 3 and 4, then all yield Gamma functions, to give:
\begin{equation}
\begin{split}
&I(a;d)=\frac{1}{\Gamma(a_1)\Gamma(a_2)\Gamma(a_5)\Gamma(d-a_{125})}\int\!\frac{\dd z_1\dd z_2}{(2\pi i)^2}\,\phi(z_1;a_3,a_{15})\,\phi(z_2;a_4,a_{25})\,\psi(z_1+z_2)\,,\\
&\phi(z;b,c)=\frac{\Gamma(-z)\,\Gamma(\tfrac d2-b+z)\,\Gamma(\tfrac d2-c-z)}{\Gamma(b-z)}\,,\\
&\psi(z)=\frac{\Gamma(a_5+z)\,\Gamma(a_{34}-\tfrac d2-z)\,\Gamma(a_{125}-\tfrac d2+z)}{\Gamma(d-a_{34}+z)}\,.
\end{split}
\label{eq:MB}
\end{equation}
The contours are straight lines parallel to the imaginary axes, on which all nine Gamma functions in the numerator have arguments with positive real part, that is,
\begin{equation}
\begin{gathered}
\re a_3-\tfrac d2<\re z_1<\min\big(0,\tfrac d2-\re a_{15}\big)\,,\\
\re a_4-\tfrac d2<\re z_2<\min\big(0,\tfrac d2-\re a_{25}\big)\,,\\
\max\big(-\re a_5,\tfrac d2-\re a_{125}\big)<\re(z_1+z_2)<\re a_{34}-\tfrac d2\,.
\end{gathered}
\label{eq:MBcond}
\end{equation}
We integrate numerically along these straight contours, along which the integrand decays exponentially~\cite{Czakon:2005rk}.

\subsection{Numerical evaluation and implementation}
\label{sec:numerics}
{Non-zero imaginary parts of the propagator powers do not affect the decay rate of the integrand in eq.~\eqref{eq:MB}, but rather change where it is largest and increase its oscillatory behavior.} {For instance, the factor $\Gamma(a_5+z_1+z_2)$ is largest near the line $\im(z_1+z_2)=-\im a_5$, which moves away from the origin as $\im a_5$ grows. A general-purpose adaptive integrator may fail to capture the dominant region of the integrand, or may resolve} its oscillations poorly. In the M.Sc.~thesis~\cite{Cappellazzi:2026msc} it was observed that adaptive evaluations of eq.~\eqref{eq:MB} lost accuracy for this reason once $|\im a_i|\gtrsim10$. There, numerical stability was restored through a suitable reparametrization of the contours. Here we propose an improved approach.

We sum the integrand over a uniform lattice of points on the straight contours, with step $h$ in both variables, and multiply by $h^2$. For a rapidly decaying integrand that is analytic in a strip of width $2\delta$ around the contour, where $\delta$ is the distance of the contour from the nearest pole, this sum differs from the integral by a term of order $e^{-2\pi\delta/h}$~\cite{Trefethen:2014}. The integrand
in eq.~\eqref{eq:MB} satisfies these conditions. We therefore place each contour so as to maximize $\delta$ and take $h$ to be a small enough fraction of $\delta$ to reach the desired precision; the lattice sum then agrees with the integral at the level of double-precision round-off, also at large imaginary parts of the powers. 

Where no straight contours exist, the integral is defined by analytic continuation, as in the Mathematica package \texttt{MB.m}~\cite{Czakon:2005rk}: one starts from propagator powers for which straight contours exist, moves the real parts of the powers along a straight line to the target values, and picks up the residue of every pole that crosses a contour along the way. The result is the two-fold integral, plus one-fold MB integrals and products of Gamma functions. The one-fold integrals are performed by summing over a lattice in the same way. The positions of the contours and the starting point of the continuation are free within the constraints of eq.~\eqref{eq:MBcond}. We choose them by linear programming so as to maximize the distance $\delta$ for the two-fold integral and all residue terms. 

\subsection{Accuracy of the kite integrals}
\label{sec:checks}

The program takes as input the five propagator powers, the dimension and a requested relative tolerance, and returns the value of the corresponding kite integral with an error estimate, refining the lattice until the estimate is below the tolerance. The estimate is the change of the sum when the lattice step is doubled, plus the round-off error of the sum.

We tested the program against the closed forms for a zero or negative integer power, the exact result of ref.~\cite{Kazakov:1983pk} for the family $I(1,1,1,1,\alpha;3)$, the integration-by-parts triangle rule~\cite{Chetyrkin:1981qh,Tkachov:1981wb} at random complex powers, the 48 symmetry images of section~\ref{sec:def}, and the hypergeometric representations of appendix~\ref{app:hyp} for the integrals with two unit powers. Reference values of much higher precision, at generic complex powers (including selected cases relevant to the practical power spectrum computation that exhibit the largest imaginary parts) are reproduced at a requested tolerance of $10^{-10}$, always within the error estimate.

\section{From the power spectrum to kite integrals}
\label{sec:pk}

\subsection{Two-loop contributions to the power spectrum}
The power spectrum $P$ measures the variance of the matter density contrast at wavenumber $k$. On large scales the fluctuations evolve linearly, {so $P$ is well} described by the linear spectrum $\plin$. On smaller scales gravity couples modes of different wavenumbers, and perturbation theory organizes the {resulting corrections} as a loop expansion, in which every loop is an integral over an {internal wavevector, or momentum, $\mathbf{q}$~\cite{Bernardeau:2001qr}.} 
In the Einstein--de~Sitter approximation for the time dependence, the power spectrum at a given redshift is
\begin{equation}
P(k)=\plin(k)+P_{\rm 1L}(k)+P_{\rm 2L}(k)+\dots\,,
\label{eq:loopexp}
\end{equation}
{where the one- and two-loop corrections are}
\begin{align}
P_{\rm 1L}&=P_{13}+P_{22}\,,
\label{eq:P1L}\\
P_{\rm 2L}&=P_{15}+P_{24}+P_{33\text{-}\rm I}+P_{33\text{-}\rm II}\,.
\label{eq:P2L}
\end{align}
{Writing $k=|\kk|$, $q=|\qq|$ and $p=|\pp|$ for the magnitudes of the momenta, and $\int_{\qq}=\int\dd^3q/(2\pi)^3$,}
\begin{equation}
\begin{split}
P_{13}(k)&=6\,\plin(k)\intq F_3(\kk,\qq,-\qq)\,\plin(q)\,,\\
P_{22}(k)&=2\intq F_2(\qq,\kk-\qq)^2\,\plin(q)\,\plin(|\kk-\qq|)\,,
\end{split}
\label{eq:oneloop}
\end{equation}
while the terms contributing at two loops read~\cite{Carrasco:2013mua}
\begin{equation}
\begin{split}
P_{15}(k)&=30\,\plin(k)\intq\intp F_5(\kk,\qq,-\qq,\pp,-\pp)\,\plin(q)\,\plin(p)\,,\\
P_{24}(k)&=24\intq\intp F_2(\qq,\kk-\qq)\,F_4(\qq,\kk-\qq,\pp,-\pp)\,\plin(q)\,\plin(|\kk-\qq|)\,\plin(p)\,,\\
P_{33\text{-}\rm I}(k)&=\frac{P_{13}(k)^2}{4\plin(k)}\,,\\
P_{33\text{-}\rm II}(k)&=6\intq\intp F_3(\qq,\pp,\kk-\qq-\pp)^2\,\plin(q)\,\plin(p)\,\plin(|\kk-\qq-\pp|)\,.
\end{split}
\label{eq:diagrams}
\end{equation}
The symmetrized perturbation-theory kernels $F_n$ are known in closed form as rational functions of the momenta, generated by a recursive relation~\cite{Goroff:1986ep,Jain:1993jh}. 

The counterterms of the EFTofLSS at this order involve at most one-loop integrals, and are discussed in appendix~\ref{sec:ctlist}.

\subsection{FFTLog decomposition}
\label{sec:fftlog}
In the FFTLog prescription~\cite{Hamilton:1999uv,Simonovic:2017mhp}, we sample the linear spectrum $\plin(k)$ at $N$ logarithmically spaced points $k_r=k_{\min}(k_{\max}/k_{\min})^{r/N}$, $r=0,\dots,N-1$, and write it as a sum of complex power laws,
\begin{equation}
\plin(k)\simeq\sum_{m=-N/2}^{N/2}c_m\,k^{\nu+i\eta_m}\,,\qquad
\eta_m=\frac{2\pi m}{\log(k_{\max}/k_{\min})}\,,
\label{eq:fftlog}
\end{equation}
with coefficients from a discrete Fourier transform,
\begin{equation}
c_m=\frac{W_m}{N}\sum_{r=0}^{N-1}\frac{\plin(k_r)}{k_r^{\nu}}\,k_{\min}^{-i\eta_m}e^{-2\pi i mr/N}\,,\qquad
W_m=\begin{cases}\tfrac12\,, & |m|=N/2\,,\\ 1\,, & |m|<N/2\,.\end{cases}
\label{eq:cm}
\end{equation}
{The auxiliary parameter $\nu$, called the bias, is a real number that can be arbitrarily chosen.}
{Importantly,} the exponents depend only on the {chosen FFTLog} grid, that is on $N$, $\nu$, $k_{\min}$ and $k_{\max}$, {while} all information about {the} cosmology sits in the coefficients $c_m$. 

A power law in momentum space has the form of a propagator of a massless particle raised to a complex power,
\begin{equation}
k^{\nu+i\eta_m}=(k^2)^{-X_m}\,,\quad X_m=-\frac{\nu+i\eta_m}{2}\,.
\label{eq:Xm}
\end{equation}
Therefore, by inserting the decomposition~\eqref{eq:fftlog} for each factor of $\plin$ in a loop integral, the latter reduces to a combination of massless Feynman integrals with complex propagator powers.

The sum of power laws, eq.~\eqref{eq:fftlog}, is defined for all $k$. Outside the sampled interval, it repeats the same pattern periodically in $\log k$, multiplied by $k^\nu$, and does not resemble the shape of the linear spectrum. We therefore split the sum into its parts below, on and above the sampled interval,
\begin{equation}
\sum_{m=-N/2}^{N/2}c_m\,k^{\nu+i\eta_m}=P^<(k)+P^{\rm cut}(k)+P^>(k)\,,
\label{eq:split}
\end{equation}
where each part equals the sum on its corresponding momentum range and vanishes elsewhere. In practice, $P^{\rm cut}$ is the actual reconstruction of the linear spectrum on the FFTLog interval; $P^<$ and $P^>$ affect the loop corrections only in the limits of very small and very large loop momenta. This is discussed in more detail in section~\ref{sec:compare}. 

Throughout the analysis presented in the following, we use the linear power spectrum of the fiducial $\Lambda$CDM cosmology of ref.~\cite{Anastasiou:2025jsy} at redshift $z=0.57$\footnote{$h=0.673$, $\omega_b=0.02237$, $\omega_{\rm cdm}=0.1203$, $\ln(10^{10}A_s)=3.044$, $n_s=0.965$, one massive neutrino with $m_\nu=0.06$\,eV.}, obtained with \texttt{CLASS}~\cite{Blas:2011rf} from $10^{-5}$ to $10^4\hmpc$, continued as $k^{-3}$ above the chosen momentum range, and as $k$ below.

\begin{figure}[t]
\centering
\includegraphics[width=0.8\textwidth]{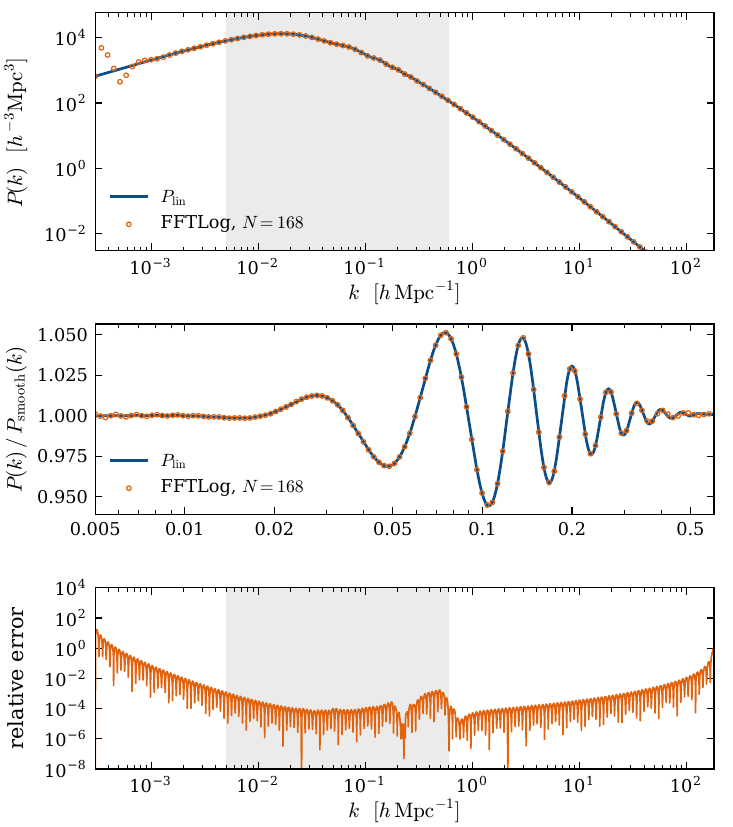}
\caption{The linear spectrum with BAO at $z=0.57$ (line) and its FFTLog reconstruction, eq.~\eqref{eq:fftlog}, with $N=168$ and $\nu=-1.6$ (circles). Top: the whole FFTLog interval; the shaded band is the range of $k$ studied in this paper. Middle: the spectrum divided by a smooth version without BAO, which serves only to make the oscillations visible, shown only in the shaded range. Bottom: absolute value of the relative error of the reconstruction, again over the whole FFTLog interval.}
\label{fig:plin}
\end{figure}

We sample it with $N=168$ points on the FFTLog interval
\begin{equation}
k_{\min}=3\times10^{-4}\hmpc\,,\quad k_{\max}=180\hmpc\,,
\label{eq:interval}
\end{equation}
with bias $\nu=-1.6$. The largest frequency is $\eta_{N/2}\approx40$, so the powers $X_m$ have imaginary parts up to about 20. Figure~\ref{fig:plin} shows the spectrum and its FFTLog reconstruction. Since the points are equally spaced in $\log k$, while the BAO have a fixed period in $k$, the relative error of the reconstruction grows from a few times $10^{-4}$ between $0.01$ and $0.3\hmpc$ to $0.2\%$ near $0.6\hmpc$. A more accurate reproduction of the BAO can be achieved by increasing $N$.  The large relative error towards the ends of the FFTLog interval is instead due to ringing artifacts arising from the periodic continuation of $\plin(k)\,k^{-\nu}$ being discontinuous~\cite{McEwen:2016fjn}. In any case, its impact on the loop computations is negligible.

\subsection{Precomputed matrices}
\label{sec:matrices}
The integrand of each two-loop contribution is a rational function of the loop momenta times factors of $\plin$. After a linear change of the loop momenta and an expansion of the rational function, each term becomes a kite integral, eq.~\eqref{eq:kite} with $d=3$, once the factors of $\plin$ are replaced by the power laws of eq.~\eqref{eq:fftlog}: each power law adds the complex power $X_m$ of eq.~\eqref{eq:Xm} to the integer power of its line. For $P_{24}$ and $P_{33\text{-}\rm II}$, which contain three spectra in the loops,
\begin{equation}
P_{24}(k)\;\text{or}\;P_{33\text{-}\rm II}(k)=\frac{k^6}{64\pi^3}\sum_{m_1,m_2,m_3}c_{m_1}c_{m_2}c_{m_3}\,k^{-2(X_{m_1}+X_{m_2}+X_{m_3})}\,M(m_1,m_2,m_3)\,,
\label{eq:tensor}
\end{equation}
where the tensor $M$, different for the two contributions, is a sum of kite integrals with rational coefficients. For $P_{15}$ the analogous formula has $\plin(k)$ times a matrix, and $P_{33\text{-}\rm I}$ needs only the one-loop vector of ref.~\cite{Simonovic:2017mhp}. The tensors depend only on the frequency grid and are computed once. Only half of their entries need to be computed: changing the signs of all $m_i$ conjugates every power, and therefore every kite integral, so that $M(-m_1,-m_2,-m_3)$ is the complex conjugate of $M(m_1,m_2,m_3)$. We call these tensors, and the matrix of $P_{15}$, the precomputed matrices; they are computed and stored separately for each contribution.

For each cosmology, eq.~\eqref{eq:tensor} contracts the precomputed tensor with the coefficients $c_m$ of the input spectrum. Since $k$ enters only through the sum of the three frequencies, we first reduce the tensor and the coefficients to about $3N/2$ numbers, one for each value of this sum, and then evaluate a single sum at every wavenumber. All kite integrals are evaluated once, for the tensors; a new cosmology costs one fast Fourier transform and the contraction, a few milliseconds.

\subsection{Classes of kite integrals}
\label{sec:classes}
Every two-loop contribution to the power spectrum contains at most three factors of $\plin$ inside the loops, each on a different line of the kite integral, and each contributes a complex power (section~\ref{sec:matrices}). The perturbation theory kernels contribute only integer powers. At least two of the five powers are therefore integers, and the integral simplifies whenever one of them is zero or negative. A zero power removes its corresponding line, and the kite integral factorizes into two one-loop bubble integrals, each a ratio of Gamma functions. A negative integer power turns the line into a polynomial in the momenta in the numerator, and the kite integral becomes a finite sum of products of bubbles~\cite{Simonovic:2017mhp}.\footnote{For a negative integer power, ref.~\cite{Simonovic:2017mhp} gave the sum in three dimensions; our program uses an equivalent sum valid in any dimension, checked against it and against the MB representation.} 

A genuine two-loop integral remains only when all integer powers are positive; in the power spectrum they then all equal one. Up to symmetry, two cases remain. The integral $I(a_1,1,a_3,a_4,1;3)$ occurs in $P_{15}$ and $P_{24}$, and $I(1,a_2,a_3,1,a_5;3)$ in $P_{33\text{-}\rm II}$; we evaluate both with eq.~\eqref{eq:MB}. For the first, the reflection gives the single-series representation of ref.~\cite{Kotikov:1995cw}; for the second, the same technique gives a double series, which one of us reduced to single sums of hypergeometric series at unit argument, eq.~\eqref{eq:hyp-opp}. Both serve as checks of our program (appendix~\ref{app:hyp}).

Of the about 4400 distinct kite integrals in the dark-matter power spectrum, more than 95\% have a zero or negative integer power. The remaining 167 configurations of powers dominate the computing time; their accurate and fast evaluation, described in section~\ref{sec:numerics}, is the main new technical contribution of this paper.

\section{The two-loop power spectrum}
\label{sec:spectrum}

\subsection{Result}
\label{sec:result}
\begin{figure}[t]
\centering
\includegraphics[width=0.74\textwidth]{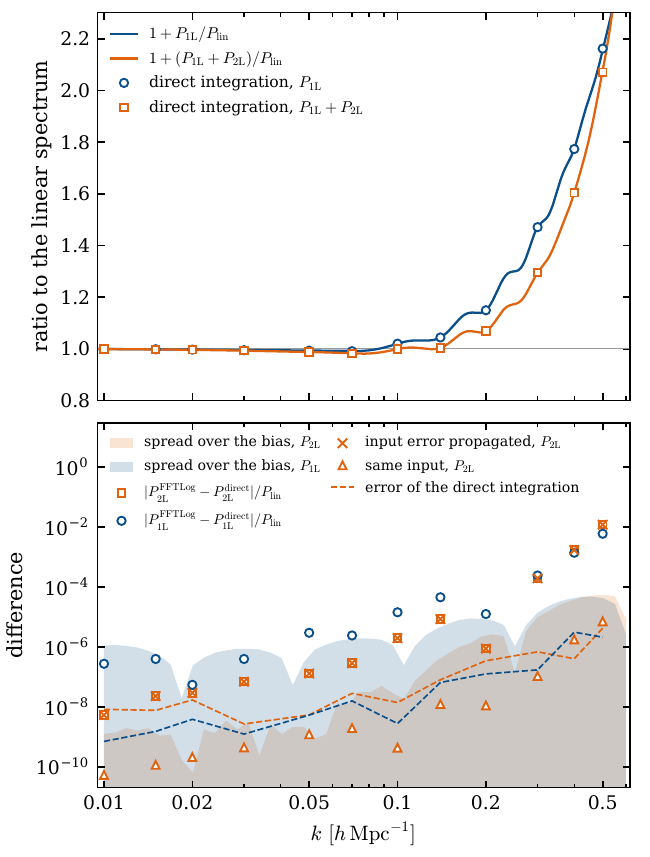}
\caption{One- and two-loop corrections to the $\Lambda$CDM power spectrum with BAO at $z=0.57$, divided by the linear spectrum, without counterterms. Lines: the FFTLog corrections $P^{\rm FFTLog}_{\rm 1L}$ and $P^{\rm FFTLog}_{\rm 1L}+P^{\rm FFTLog}_{\rm 2L}$ of eq.~\eqref{eq:fftlogdef}, for $N=168$ and $\nu=-1.6$. Circles and squares: the direct numerical integration of eqs.~\eqref{eq:oneloop} and \eqref{eq:diagrams} with the physical linear spectrum over all loop momenta (section~\ref{sec:compare}), at 12 wavenumbers, for one loop and for one plus two loops. Bottom: the difference between the FFTLog corrections and the direct integration at one loop (circles) and two loops (squares), the error of the direct integration (dashed), and the difference between the FFTLog corrections computed with $\nu=-1.6$ and with $\nu=-1.8$, each with its own terms $\Delta$ (bands). Crosses: the effect of the reconstruction error of the input on the two-loop integral, $|P^{\rm cut}_{\rm 2L}-P^{\rm phys}_{\rm 2L}|/\plin$ (section~\ref{sec:compare}), which coincides with the squares. Triangles: the remainder of the two-loop comparison when the direct integration is given the reconstructed spectrum as input (section~\ref{sec:compare}), with no free parameter.}
\label{fig:spectrum}
\end{figure}
Figure~\ref{fig:spectrum} shows the one- and two-loop corrections for the spectrum with BAO. The one-loop correction reaches about $15\%$ of the linear spectrum at $0.2\hmpc$ and exceeds it above $0.45\hmpc$. 
The two-loop correction is negative, below one percent of the linear spectrum up to $0.1\hmpc$, and reaches a fifth of it between $0.3$ and $0.4\hmpc$; the full spectrum through two loops is about twice the linear one at $0.5\hmpc$. The steps in the ratios are the BAO: large-scale displacements damp them, and this damping is usually resummed~\cite{Senatore:2014via,Baldauf:2015xfa}; we do not resum it, so the loop corrections contain the oscillating part of the linear spectrum with coefficients that grow with $k$.

The corrections are shown divided by the linear spectrum, which removes its overall shape from the plot and shows their relative size directly. The lines are the FFTLog corrections of eq.~\eqref{eq:fftlogdef} and the circles and squares the direct integration of the physical loop integrals; the two calculations share nothing but the input spectrum. The next section describes the direct integration and the comparison. In short, the two agree with no free parameter, at one loop to $5\times10^{-5}$ of the linear spectrum up to $0.2\hmpc$ and at two loops to $10^{-5}$; the remaining difference, at all wavenumbers, is the accuracy with which the power laws represent the input spectrum.

\subsection{Comparison with the direct integration}
\label{sec:compare}
Figure~\ref{fig:spectrum} shows, as circles and squares, a direct numerical integration of eqs.~\eqref{eq:oneloop} and \eqref{eq:diagrams} over all loop momenta with the \texttt{CLASS} linear power spectrum of section~\ref{sec:fftlog}; this gives the physical loop integrals, $P^{\rm phys}_{\rm 1L}$ and $P^{\rm phys}_{\rm 2L}$. The integration runs over the lengths and directions of the loop momenta, with a smooth partition of unity following refs.~\cite{Carrasco:2013sva,Blas:2013aba}, so that every momentum that can become small is itself an integration variable and the integrable infrared enhancements are resolved. It covers loop momenta between $10^{-4}$ and $180\hmpc$; the contributions from outside this range are computed from the limits of the integrands at small and large loop momenta and added (appendix~\ref{app:ctuv}). The error of the points, estimated from two settings of the integration, is below $5\times10^{-6}$ of the linear spectrum.

The FFTLog calculation gives first what we call the naive corrections, $P^{\rm naive}_{\rm 1L}$ and $P^{\rm naive}_{\rm 2L}$: the same loop integrals with $\plin$ replaced by the whole sum of power laws, eq.~\eqref{eq:fftlog}, and each power law integrated over all loop momenta in analytic regularization. Each such integral is a massless Feynman integral with complex powers, defined by analytic continuation in the powers~\cite{Simonovic:2017mhp}: where it converges it is the ordinary integral, and where it diverges as a power of a cutoff the continuation sets the divergent part to zero, as dimensional regularization does~\cite{Scoccimarro:1996se,Pajer:2013jj}. In practice $P^{\rm naive}$ is the contraction of the precomputed tensors with the coefficients $c_m$, eq.~\eqref{eq:tensor}.

The naive corrections differ from the physical loop integrals in two ways. Both come from loop momenta far below or far above $k$, where the integrands take universal forms~\cite{Goroff:1986ep,Jain:1995kx,Scoccimarro:1995if,Carrasco:2013mua,Baldauf:2015aha}. Both are computed in appendix~\ref{app:ctuv}, eqs.~\eqref{eq:ct1} and \eqref{eq:ct}; here we summarize them.
\begin{enumerate}
\item \emph{Loop momenta outside the FFTLog interval}, where the power laws $P^<$ and $P^>$ differ from the linear spectrum. Loop momenta far below $k$ enter only through the variance of $P^<$, since the uniform displacement they produce cancels at each loop order~\cite{Jain:1995kx,Scoccimarro:1995if}, and this variance is tiny for our bias. One loop momentum far above $k$ gives a moment of the spectrum above the interval, a constant, times a one-loop integral in which one kernel is replaced by its large-momentum limit~\cite{Simonovic:2017mhp}; only the difference between the moment of the power laws and that of the linear spectrum enters.
\item \emph{Both loop momenta large.} The integrals of the power laws converge here only for $\nu<-2$ (section~\ref{sec:bias}). For our bias, analytic regularization sets this double-hard term, which is scaleless for a power law, to zero, whereas for the linear spectrum it is a constant times $k^2\plin$. We compute this constant from the large-momentum limit of the integrand of $P_{15}$ by the method of regions~\cite{Beneke:1997zp} (appendix~\ref{app:ctuv}); the analogous terms of order $k^4$ are negligible for $\nu<-1$.
\end{enumerate}
All these terms are computed from the coefficients $c_m$ and the linear spectrum, with no free parameter. Their forms, $k^2\plin$, $k^4\plin$, $k^4$ and the one-loop integrals of eq.~\eqref{eq:shapes}, are exactly those of the counterterms of the EFTofLSS at two loops~\cite{Carrasco:2013mua,Anastasiou:2025jsy,Bakx:2026rmd} (appendix~\ref{sec:ctlist}). This is no accident: loop momenta above the scale where perturbation theory fails contribute to the physical loop integrals as well, and the counterterms exist to absorb this dependence~\cite{Baumann:2010tm,Carrasco:2012cv}. The renormalized spectrum, loop corrections plus counterterms with fitted coefficients, is therefore the same whether the loop integrals are computed with a cutoff, with the physical spectrum to infinite momenta or in analytic regularization; only the fitted coefficients change.

We remove both effects and define the FFTLog corrections as
\begin{equation}
P^{\rm FFTLog}_{\rm 1L}=P^{\rm naive}_{\rm 1L}-\Delta_{\rm 1L}\,,\qquad P^{\rm FFTLog}_{\rm 2L}=P^{\rm naive}_{\rm 2L}-\Delta_{\rm 2L}\,,
\label{eq:fftlogdef}
\end{equation}
where $\Delta_{\rm 1L}$ and $\Delta_{\rm 2L}$ are the right-hand sides of eqs.~\eqref{eq:ct1} and \eqref{eq:ct}. These are the lines of figure~\ref{fig:spectrum}. They should agree with the physical loop integrals without any further counterterm, up to the reconstruction error of the input spectrum on the interval, and they should not depend on the bias (section~\ref{sec:bias}).

\begin{table}[t]
\centering
\begin{tabular}{lcccc}
\toprule
& \multicolumn{2}{c}{one loop} & \multicolumn{2}{c}{two loops}\\
input of the direct integration & $k\le0.2$ & $k\le0.5$ & $k\le0.2$ & $k\le0.5$\\
\midrule
linear spectrum & $5\times10^{-5}$ & $6\times10^{-3}$ & $9\times10^{-6}$ & $1.2\times10^{-2}$\\
reconstructed spectrum $P^{\rm cut}$ & $7\times10^{-8}$ & $2\times10^{-6}$ & $1\times10^{-8}$ & $7\times10^{-6}$\\
\midrule
spread between $\nu=-1.6$ and $-1.8$ & $9\times10^{-6}$ & $5\times10^{-5}$ & $2\times10^{-6}$ & $6\times10^{-5}$\\
\bottomrule
\end{tabular}
\caption{Largest difference between the FFTLog corrections and the direct integration, in units of the linear spectrum, for $k$ up to $0.2$ and up to $0.5\hmpc$, with the linear spectrum or its reconstruction as input of the direct integration; last row, the spread between two biases.}
\label{tab:cmp}
\end{table}

The lower panel of figure~\ref{fig:spectrum} and table~\ref{tab:cmp} give the difference between the FFTLog corrections and the direct integration. Up to $0.2\hmpc$ the two agree to better than $10^{-4}$ of the linear spectrum at both orders; towards $0.5\hmpc$ the difference grows to about a percent, far above the error of the direct integration. What remains is the error of the input: on the interval the FFTLog calculation integrates $P^{\rm cut}$ and the direct integration $\plin$, which differ by the reconstruction error of figure~\ref{fig:plin}. At two loops this is seen directly: the crosses in the lower panel, the two-loop integral of $P^{\rm cut}$ minus that of $\plin$, both by direct integration, coincide with the squares at every wavenumber. Below $0.2\hmpc$ the difference comes from the ringing of the reconstruction near $k_{\max}$, which enters through the moments of the spectrum at large momenta; above $0.3\hmpc$ the reconstruction error in the BAO range takes over, amplified by the loop integrals.

We verified this by giving the direct integration the same input, $P^{\rm cut}$ in place of $\plin$. The result differs from the naive corrections only through the loop momenta outside the interval, so the terms of eqs.~\eqref{eq:ct1} and \eqref{eq:ct} apply with the full moment of the power laws in place of the difference, and with the one-loop integrals and the constants evaluated for $P^{\rm cut}$. With these terms the two calculations agree to the precision of the direct integration (triangles in figure~\ref{fig:spectrum}, second row of table~\ref{tab:cmp}), again with no free parameter, also at $\nu=-0.8$, where the moment term is a hundred times the spectrum at $0.5\hmpc$ and the terms of order $k^4$ are needed. The FFTLog calculation is therefore exact for the spectrum it is given, and the computed terms are correct to the same level. These terms are not small: at $\nu=-1.6$ the double-hard term is half the linear spectrum at $0.5\hmpc$.

We repeated this test for a second $\Lambda$CDM cosmology with the same tensors.\footnote{$h=0.70$, $\omega_b=0.0225$, $\omega_{\rm cdm}=0.135$, $\ln(10^{10}A_s)=2.95$, $n_s=0.94$, one massive neutrino with $m_\nu=0.24$\,eV, total-matter spectrum at $z=0.57$ from \texttt{CLASS}, with $\sigma_8=0.80$ at $z=0$.} The computed terms account for the difference at the same level, with no free parameter, as expected since the tensors do not depend on the input spectrum.

\subsection{Choice and independence of the bias}
\label{sec:bias}
For a single power law, the one-loop sum converges for $-3<\nu<-1$~\cite{Pajer:2013jj,Simonovic:2017mhp}: the infrared condition is that of the density variance, and the ultraviolet one that of one large loop momentum in $P_{13}$, which behaves as $k^2\plin(k)\int\dd q\,q^{\nu}$. FAST-PT requires the same range and uses $\nu=-2$~\cite{McEwen:2016fjn}. At two loops the regions with one large loop momentum again require $\nu<-1$, while two large loop momenta give the term $k^2\plin(k)\int\dd q\,q^{2\nu+3}$ in $P_{15}$, which converges only for $\nu<-2$~\cite{Carrasco:2013mua}. For $-2<\nu<-1$ the analytic continuation therefore drops exactly the term we compute. We checked this with a grid of $N=256$ power laws at $\nu=-2.2$, inside the window of convergence, with the input windowed near the ends of the interval as in ref.~\cite{McEwen:2016fjn}: there the formula of appendix~\ref{app:ctuv} gives a coefficient smaller by a factor of 200, and a fit to the direct integration reproduces it to $1\%$.

We use $\nu=-1.6$, for two reasons. First, the terms of eq.~\eqref{eq:fftlogdef} are small there: towards $\nu=-1$ the moment of the power laws above the interval diverges, and with it the amplification of the input error; at $\nu=-0.8$, the moment term reaches 100 times the linear spectrum at $0.5\hmpc$ and the agreement with the direct integration degrades above $0.3\hmpc$. Towards $\nu=-3$ the variance of the power laws below the interval grows instead. Second, inside the window of convergence the periodic continuation is strongly discontinuous and the reconstruction error at low $k$ grows by an order of magnitude, while nothing is gained for the renormalized spectrum, since the dropped term has the form of the leading counterterm.

The bias changes the power laws outside the interval, the dropped term and, through the reconstruction error, slightly the spectrum on the interval. The FFTLog corrections of eq.~\eqref{eq:fftlogdef}, each bias with its own computed terms, must therefore agree between biases. The bands in figure~\ref{fig:spectrum} show the spread between $\nu=-1.6$ and $-1.8$ (last row of table~\ref{tab:cmp}): it is smaller than the difference to the direct integration, which both share, while the computed double-hard terms differ by $5\%$, as they must. The bias affects the result only through the reconstruction error.


\subsection{Accuracy and computational cost}
\label{sec:validation}
The precision of the two-loop spectrum is set by the representation of the input spectrum, not by the kite integrals. The error estimates of the kite integrals, propagated through the tensors, put their effect on $P_{\rm 2L}$ far below the levels discussed here, and the comparisons to direct numerical integration find no difference beyond the reconstruction error discussed in section~\ref{sec:fftlog}. The loop corrections are multilinear in the spectrum and amplify it, at $0.5\hmpc$ by a factor of about four at one loop and eight at two loops (section~\ref{sec:compare}), and the terms of counterterm form carry the same relative error.
Repeating the calculation with 256 power laws, whose reconstruction error is much smaller, changes the FFTLog corrections of eq.~\eqref{eq:fftlogdef} by about $10^{-4}$ of the linear spectrum up to $0.2\hmpc$, and by about one percent of it between $0.3$ and $0.6\hmpc$ at two loops. 
Most of the latter is the reconstruction error of the external factor $\plin(k)$ multiplied by the displacement by the long modes, the terms of appendix~\ref{sec:allchecks}; 
with the external factor taken from the input spectrum and these terms included, the change is a few parts in $10^{3}$ of the linear spectrum. The comparison with the published two-loop spectrum of ref.~\cite{Anastasiou:2025jsy} is made in this form and agrees at this level (appendix~\ref{sec:lit}). A second $\Lambda$CDM cosmology gives the same agreement with a direct integration as the fiducial one (section~\ref{sec:compare}). Since the matrices do not depend on the input spectrum, these tests carry over to any spectrum the grid resolves.

The cost lies almost entirely in the precomputation, and there in the 167 configurations of powers that require the MB representation (section~\ref{sec:classes}): those with opposite unit lines and those with adjacent unit lines and three complex powers take about half each, all other integrals together less than one percent. The tensors for $N=168$ and $\nu=-1.6$ took about 400 core-hours with a Python program on x86 nodes of the Max Planck Computing and Data Facility; they occupy 13\,MB. A new cosmology then takes about 6\,ms for 100 wavenumbers and 17\,ms for 1000, on one core and including the fast Fourier transform of the input spectrum.

\section{Discussion and conclusions}
\label{sec:discussion}

We have constructed cosmology-independent matrices for the two-loop matter power spectrum using FFTLog and kite integrals, on a grid that resolves the BAO features of the linear spectrum. These matrices allow rapid evaluation of the two-loop power spectrum for different cosmologies. We have also established the relation between the analytically regularized FFTLog calculation and the physical loop integrals. Their difference consists of terms with the same form as the counterterms of the EFTofLSS, arising from loop momenta outside the sampled interval and from the double-hard region omitted by analytic regularization for our choice of bias. We compute these terms directly from moments of the spectrum and the large-momentum limits of the integrands, without introducing free parameters. 
Once they are included, the FFTLog spectrum agrees with the direct integration to better than $10^{-4}$ of the spectrum up to $0.2\hmpc$, and to a few parts in $10^{8}$ when both are given the same input (section~\ref{sec:compare}).
The remaining difference, about a percent of the linear spectrum at $0.5\hmpc$, is the reconstruction error of the input, $2\times10^{-3}$, amplified by the loops. 
We find the same level of agreement for two other biases and a second cosmology. After converting between subtraction schemes, our result also agrees with ref.~\cite{Anastasiou:2025jsy} to about $0.1\%$ up to $0.6\hmpc$ (appendix~\ref{sec:lit}). The matrices need to be computed only once, after which evaluating a new cosmology takes a few milliseconds.

For the kite integral, the main numerical insight is that large imaginary parts of the powers do not slow the decay of the MB integrand, but only shift the region in which it is large. A uniform lattice can therefore evaluate the integral reliably without locating this region in advance (section~\ref{sec:numerics}). The error decreases exponentially as the lattice is refined, with an error estimate available at every point. Independent hypergeometric representations confirm the implementation for configurations with two unit powers (appendix~\ref{app:hyp}). We expect the same approach to be useful for other MB integrals involving complex powers.

Direct numerical integration requires no precomputation and includes exact time dependence~\cite{Garny:2022fsh}, redshift space~\cite{Taule:2023izt} and biased tracers~\cite{Ivanov:2026zos}, but must be repeated for each cosmology, at 10 to 100\,s per wavenumber for a Monte Carlo precision of $10^{-3}$~\cite{Bakx:2025jwa}. Ref.~\cite{Anastasiou:2025jsy} instead expands the linear spectrum around a reference cosmology and precomputes the integrals by Monte Carlo. The closest method to ours is the two-loop extension of COBRA~\cite{Bakx:2025jwa}, which represents the linear spectrum by about ten functions adapted to a range of $\Lambda$CDM parameters, precomputes their two-loop integrals by Monte Carlo integration on a fixed set of wavenumbers, and evaluates a new cosmology in about a millisecond, to about $0.1\%$ within that range. Our tensors require more precomputation, but apply to any linear spectrum resolved by the grid, at any $k$ well inside the sampled interval, with about ten digits of accuracy in each kite integral. The two approaches are complementary: the COBRA basis functions can be decomposed into power laws on the same grid, allowing their two-loop templates to be precomputed by contracting our tensors rather than by Monte Carlo integration, as has been done for the one-loop bispectrum~\cite{Bakx:2025pop}. This combination could be particularly useful at higher loop orders, for instance at three loops, where the cost of direct template integration becomes increasingly large. Likewise, the matter part of the redshift-space calculation of ref.~\cite{Philcox:2026myi} could be obtained from our tensors.

Infrared resummation of the BAO~\cite{Senatore:2014via,Baldauf:2015xfa,Blas:2016sfa,Lewandowski:2018ywf} requires no new tensors. It replaces the linear spectrum inside the loops by its smooth and oscillating parts, with damping factors that depend only on the external wavenumber. Since the loop corrections are multilinear in the spectrum, the resummed result can be obtained by contracting the same tensors with the FFTLog coefficients of these parts. Natural extensions include biased tracers~\cite{Bakx:2025cvu,Bakx:2026rmd} and redshift space, which require new tensors but no new integration technique, and the two-loop bispectrum~\cite{Baldauf:2014qfa,Baldauf:2021zlt}. More generally, separating the loop integration from the cosmological input and using MB representation of loop integrals provides a reusable basis for higher-order perturbative calculations.

\section*{Code and data availability}
The ancillary files of the arXiv submission of this article contain the program that computes the kite integrals and the precomputed matrices of eq.~\eqref{eq:tensor} for $N=168$ and $\nu=-1.6$.

\acknowledgments
We thank Giovanni Cabass, Mathias Garny, Henrique Rubira, Fabian Schmidt, Zvonimir Vlah, Denis Werth and Matias Zaldarriaga for discussions and encouragement. We acknowledge support from the Max Planck--IAS--NTU Center for Particle Physics, Cosmology and Geometry. The AI assistant Claude (Anthropic), used through Claude Code, assisted in optimizing and automating the original code of PC, in running numerical checks, and in drafting and editing the manuscript. The authors have verified all results and take full responsibility for the content of this paper.

\appendix
\section{Check against hypergeometric representations}
\label{app:hyp}
For the two classes of integrals in section~\ref{sec:classes}, the kite integral can also be written as hypergeometric series which follow from the Gegenbauer polynomial technique~\cite{Chetyrkin:1980pr,Kotikov:1995cw}. They are independent of eq.~\eqref{eq:MB}, so we used them to check our baseline code on the frequency grid. With $\tilde a\equiv\tfrac32-a$, a permutation followed by the reflection, and a permutation followed by a uniqueness relation, give
\begin{equation}
\begin{split}
&I(a_1,1,a_3,a_4,1;3)=\frac{4\pi^4}{(2\pi)^9}\,f(a_1)f(a_3)f(a_4)f(\tfrac52-a_{134})\;I(\tilde a_3,\tfrac12,a_{134}-1,\tfrac12,\tilde a_1;3)\,,\\
&f(a)=\frac{\pi^{3/2}\,\Gamma(\tilde a)}{2^{2a-3}\,\Gamma(a)}\,,\\
&I(1,a_2,a_3,1,a_5;3)=\frac{\sqrt\pi\,\Gamma(\tilde a_3)\Gamma(\tilde a_5)\Gamma(a_{35}-\tfrac12)}{\Gamma(a_3)\Gamma(a_5)\Gamma(2-a_{35})}\;I(\tilde a_3,\tfrac12,a_2,1,a_{35}-\tfrac12;3)\,.
\end{split}
\label{eq:hyp-maps}
\end{equation}
Other permutations give further relations of the same kind. 

For the integral on the right of the first line,
\begin{equation}
\begin{split}
I(c_1,\tfrac12,c_3,\tfrac12,c_5;3)={}&\frac{\Gamma(\tilde c_5)}{\sqrt\pi\,\Gamma(c_5)\,(\tfrac52-c_{135})}\Big[T(c_1,c_3)+T(c_3,c_1)-R\Big]\,,\\
T(x,y)={}&\frac{\tan\pi x}{(\tfrac32-c_5)(2-c_{135})\tan\pi c_5}\,\frac{\Gamma(x+c_5-1)\,\Gamma(1-x)}{\Gamma(\tfrac52-x-c_5)\,\Gamma(x-\tfrac12)}\\
&\times{}_3F_2\!\left(1,1,y;c_{135}-1,\tfrac52-c_5;1\right),\\
R={}&\frac{\pi^3\sin\pi c_{13}}{\cos\pi c_1\,\cos\pi c_3\,\sin\pi c_{135}\,\sin\pi c_5}\\
&\times\frac{1}{\Gamma(c_1)\Gamma(c_3)\Gamma(\tfrac52-c_{35})\Gamma(\tfrac52-c_{15})}\,.
\end{split}
\label{eq:hyp-adj}
\end{equation}
The direct series  for ${}_3F_2(1,1,y;b_1,b_2;1)$ converges for $\re (2+y) < \re (b_1+b_2)$, so eq.~\eqref{eq:hyp-adj} requires $\re c_1>\tfrac12$ and $\re c_3>\tfrac12$. 
For other values of the parameters, the hypergeometric functions can always be computed by analytic continuation.

The integral on the right of the second line in eq.~\eqref{eq:hyp-maps}, $I(c_1,\tfrac12,c_3,1,c_5;3)$, has no known representation by a single series. The Gegenbauer polynomial technique of Kotikov~\cite{Kotikov:1995cw}  
 allows one to write it as the double series
\begin{equation}
\begin{split}
I(c_1,\tfrac12,c_3,1,c_5;3)={}&\frac{1}{\sqrt\pi}\sum_{n,m=0}^{\infty}\Bigg\{\frac{\pi}{\Gamma(c_5)\,\Gamma(c_5-\tfrac12)}\,\frac{\Gamma(n+m+c_5)\,\Gamma(m+c_5-\tfrac12)}{m!\,\Gamma(n+m+\tfrac32)}\\
&\quad\times\bigg[\frac{\Gamma(n+m+c_{35}-\tfrac12)\,\Gamma(\tfrac32-c_{35}-m)}{(n+m+\tilde c_1)\,\Gamma(n+m+c_{35})\,\Gamma(2-c_{35}-m)}\\
&\quad+\frac{\Gamma(n+m+\tilde c_3)\,\Gamma(c_3-\tfrac12-m)}{(n+m+c_{15}-1)\,\Gamma(n+m+2-c_3)\,\Gamma(c_3-m)}\bigg]\\
&\quad+\frac{\Gamma(\tilde c_5)}{\Gamma(c_5)}\,\frac{\Gamma(n+m+1)\,\Gamma(m+\tfrac12)}{m!\,\Gamma(n+m+\tfrac32)}\\
&\quad\times\bigg[\frac{\Gamma(n+m+\tilde c_3)\,\Gamma(c_{35}-\tfrac32-m)}{(n+m+3-c_{135})\,\Gamma(n+m+3-c_{35})\,\Gamma(c_3-m)}\\
&\quad+\frac{\Gamma(n+m+c_{35}-\tfrac12)\,\Gamma(\tfrac12-c_3-m)}{(n+m+c_{135}-\tfrac32)\,\Gamma(n+m+1+c_3)\,\Gamma(2-c_{35}-m)}\bigg]\Bigg\}\,.
\end{split}
\label{eq:hyp-double}
\end{equation}
By rearranging the sums and applying Thomae transformations of ${}_3F_2(1)$, it is possible to reduce this double series to a single function and its reflected copy,
\begin{equation}
I(c_1,\tfrac12,c_3,1,c_5;3)=H(c_1,c_3,c_5)+H(\tfrac52-c_{15},\,2-c_{35},\,c_5)\,.
\label{eq:hyp-opp}
\end{equation}
The reflection $(c_1,c_3,c_5)\mapsto(\tfrac52-c_{15},2-c_{35},c_5)$ is an involution. Here
\begin{equation}
H(c_1,c_3,c_5)=U+V+\frac{4^{c_5-1}}{(1-c_5)\,\Gamma(2c_5-1)}\sum_{n=0}^\infty(A_n+C_n)\,,
\label{eq:hyp-H}
\end{equation}
with the two closed hypergeometric contributions
\begin{equation}
\begin{split}
U={}&-\frac{4^{c_3}\,\Gamma(2-2c_3)\,\Gamma(1-c_5)\,\sin\pi c_3}{\cos\pi c_{35}\,(c_{135}-3)\,\Gamma(2-c_{35})\,\Gamma(3-c_{35})\,\Gamma(c_5)}\\
&\quad\times{}_4F_3\!\left(1,1,\tilde c_3,3-c_{135};\tfrac32,3-c_{35},4-c_{135};1\right),\\
V={}&\frac{2\sqrt\pi\,\Gamma(\tilde c_3)\,\tan\pi c_3}{\sin\pi c_5\,(c_3-1)(c_{15}-1)\,\Gamma(2-c_{35})\,\Gamma(c_5)}\\
&\quad\times{}_4F_3\!\left(1,\tilde c_3,c_5,c_{15}-1;\tfrac32,2-c_3,c_{15};1\right),
\end{split}
\label{eq:hyp-UV}
\end{equation}
and the two residual summands
\begin{equation}
\begin{split}
A_n={}&-\frac{\tan\pi c_{35}}{n+\tilde c_1}\,\frac{\Gamma(n+c_{35}-\tfrac12)\,\Gamma(n+c_5)}{\Gamma(n+\tfrac32)\,\Gamma(n+1+c_3)}\\
&\quad\times{}_3F_2\!\left(c_3,1-c_5,\tilde c_5;n+1+c_3,2-c_5;1\right),\\
C_n={}&\frac{\sin\pi c_3}{\cos\pi c_{35}\,\cos\pi c_5}\,\frac{1}{n+3-c_{135}}\,\frac{\Gamma(n+\tilde c_3)\,\Gamma(n+1)}{\Gamma(n+3-c_{35})\,\Gamma(n+\tfrac52-c_5)}\\
&\quad\times{}_3F_2\!\left(c_3,1-c_5,\tilde c_5;2-c_5,n+\tfrac52-c_5;1\right).
\end{split}
\label{eq:hyp-AC}
\end{equation}
This form is useful for two reasons. First, the ${}_3F_2$ functions in $A_n$ and $C_n$ tend to one as $n\to\infty$. This gives the conditions for convergence. The Gamma-function prefactors behave as $n^{2\re c_5-4}$, so the series converge absolutely for $\re c_5<\tfrac32$, independently of the other parameters. This can be always achieved using symmetry transformations of eq.~\eqref{eq:hyp-maps}. Second, the ${}_3F_2$ functions can be easily computed using recursion relations for any $n$. We checked eq.~\eqref{eq:hyp-opp} against eq.~\eqref{eq:MB} and against the double series at random real and complex points to $10^{-14}$.

\section{Terms of counterterm form and comparison with published results}
\label{app:lit}
This appendix gives the terms by which the naive FFTLog corrections of section~\ref{sec:compare} differ from the physical loop integrals, relates them to the counterterms of the EFTofLSS, and compares our result with the published values of ref.~\cite{Anastasiou:2025jsy}. The renormalized spectrum is
\begin{equation}
P^{\rm ren}(k)=\plin(k)+P_{\rm 1L}(k)+P_{\rm 2L}(k)+P_{\rm ct}(k)\,,
\label{eq:Pren}
\end{equation}
where $P_{\rm ct}$ contains the counterterms. The loop corrections depend on how the loop integrals are regularized; $P^{\rm ren}$ does not, since a change of regularization shifts them by terms of the form of the counterterms, which only changes the fitted values of their coefficients. We take the counterterms from the complete two-loop renormalization of ref.~\cite{Anastasiou:2025jsy}; ref.~\cite{Carrasco:2013mua} gave an earlier treatment with fewer counterterms.

\subsection{Loop momenta outside the FFTLog interval}
\label{app:ctuv}

We denote by $\tilde P$ the sum of power laws in eq.~\eqref{eq:fftlog}, with exponents $a_m=\nu+i\eta_m$. We wish to compare two calculations of the loop corrections: the naive FFTLog calculation, in which $\tilde P$ is integrated over all loop momenta using analytic regularization, and the physical calculation, in which the input spectrum is $\plin$. Since the loop corrections are multilinear in the input spectrum, their difference can be organized according to the momentum regions in which the two prescriptions differ.

Contributions in which all loop momenta lie within the FFTLog interval cancel between the two calculations. Below $k_{\min}$, the power-law contributions enter only through their variance $\sigma^2_<=\int_{p<k_{\min}}\dd^3p\,(2\pi)^{-3}P^<(p)$, which gives a negligible correction for the parameters considered here. The remaining difference therefore comes from configurations in which at least one loop momentum exceeds $k_{\max}\gg k$.

We evaluate these high-momentum contributions using the method of regions~\cite{Beneke:1997zp}. At leading power in $k/k_{\max}$, two types of configurations contribute: a single hard loop momentum, with the remaining momenta finite, and two hard loop momenta of comparable magnitude (three in $P_{33\text{-}\rm II}$). These regions can overlap, and their common contribution must be subtracted to avoid double counting.

Consider first the single-hard region, where one loop momentum $p$ is much larger than the external momentum and the remaining loop momenta. At leading order, the angular average of the kernel takes the form $G/p^2$, where $G$ is a rational function of the remaining momenta proportional to $k^2$~\cite{Goroff:1986ep,Peebles:1980yev}. The integral over the hard momentum then factorizes into the high-momentum moment
\begin{equation}
S_{\rm UV}=\int_{p>k_{\max}}\frac{\dd^3p}{(2\pi)^3}\,\frac{\tilde P(p)}{p^2}=-\frac{1}{2\pi^2}\sum_{m=-N/2}^{N/2}\frac{c_m\,k_{\max}^{a_m+1}}{a_m+1}\,,
\label{eq:seff}
\end{equation}
multiplying a one-loop integral in which the original kernel is replaced by $G$. We refer to this remaining integral as the shape of the single-hard contribution. For $\nu<-1$, the integral defining $S_{\rm UV}$ converges and agrees with the sum above, whereas for $\nu>-1$ the sum provides its analytic continuation. The corresponding moment of the physical spectrum is denoted by $S^{\rm phys}_{\rm UV}$, and we define $\Delta S=S_{\rm UV}-S^{\rm phys}_{\rm UV}$.

The double-hard region has a different structure. When both loop momenta are hard, the kernel can be expanded in the external momentum $k$. Each term in this expansion is homogeneous in the loop momenta. Its integral against power laws over the full momentum domain is therefore scaleless and vanishes in analytic regularization. Consequently, the contribution of the naive FFTLog calculation from outside the box $q,p<k_{\max}$ is minus the integral of the expanded kernel over the box. The latter is finite and can be written as a sum over the FFTLog coefficients $c_m$. The physical calculation, by contrast, contains a nonzero high-momentum tail, which must be evaluated using $\plin$.

Combining the single-hard and double-hard contributions, with the appropriate overlap subtraction, the difference for a contribution $I$ with kernel $K$ can be expressed as
\begin{equation}
I[\tilde P]-I[\plin]=S_{\rm UV}\,\Sigma[P^{\rm cut}]-S^{\rm phys}_{\rm UV}\,\Sigma[\plin]-H[\tilde P]-2\,O[\tilde P]-R[\plin]\,,
\label{eq:regions}
\end{equation}
up to the response terms and corrections suppressed by powers of $k/k_{\max}$ and $k_{\min}/k$. Here $\Sigma$ denotes the single-hard shape, with its remaining loop momentum restricted to the FFTLog interval. The term $H$ is the integral of the double-hard expansion of $K$ over the box, while $O$ accounts for its overlap with the single-hard expansion in the region $q<k_{\max}<p$. The factor of two accounts for the two possible orderings of the hard momenta. Finally, $R$ denotes the double-hard part of the physical integral above $k_{\max}$, which converges because $\plin$ falls as $p^{-3}$ at large momenta. The reconstruction error associated with the difference between $\Sigma[P^{\rm cut}]$ and $\Sigma[\plin]$, multiplied by $S^{\rm phys}_{\rm UV}$, is included in the constants introduced below.

Summing over all contributions, we obtain the difference between the naive and physical one-loop corrections,
\begin{equation}
P^{\rm naive}_{\rm 1L}-P^{\rm phys}_{\rm 1L}=-\frac{61}{315}\,\Delta S\,k^2\plin(k)+\sigma^2_<\Big[\tfrac{2519}{2205}\plin-\tfrac{47}{105}k\plin'+\tfrac{1}{10}k^2\plin''\Big]\,,
\label{eq:ct1}
\end{equation}
and, at two loops,
\begin{equation}
\begin{split}
P^{\rm naive}_{\rm 2L}-P^{\rm phys}_{\rm 2L}={}&\Delta S\big[S_{15}(k)+S_{24}(k)+S_{33}(k)\big]+\sigma^2_<X(k)\\
&+\alpha k^2\plin(k)+\gamma k^4\plin(k)+\epsilon k^4\,,
\end{split}
\label{eq:ct}
\end{equation}
where the single-hard shapes are
\begin{equation}
\begin{split}
S_{15}(k)&=60\,\plin(k)\intq G_3(\kk,\qq,-\qq)\,P^{\rm cut}(q)\,,\qquad S_{33}(k)=-\tfrac{61}{630}\,k^2P^{\rm cut}_{13}(k)\,,\\
S_{24}(k)&=24\intq F_2(\qq,\kk-\qq)\,G_2(\qq,\kk-\qq)\,P^{\rm cut}(q)\,P^{\rm cut}(|\kk-\qq|)\,,
\end{split}
\label{eq:shapes}
\end{equation}
with $P^{\rm cut}_{13}$ denoting $P_{13}$ evaluated using $P^{\rm cut}$.

These expressions separate the high-momentum discrepancy into two parts. The terms proportional to $\Delta S$ retain a nontrivial dependence on $k$ through the one-loop shapes, whereas the remaining high-momentum contributions are analytic in the external momentum and can be collected into counterterm-like structures. At two loops, their coefficients are
\begin{equation}
\begin{split}
\alpha&=\alpha_{\rm HH}+A_0\,S_{\rm UV}\,U-r_2\,,\\
\gamma&=\gamma_{\rm HH}+A_1\,S_{\rm UV}^2+\tfrac14\big(\tfrac{61}{315}\big)^2\big[S_{\rm UV}^2-(S^{\rm phys}_{\rm UV})^2\big]\,,\\
\epsilon&=g\,S_{\rm UV}\,W-\epsilon_{24}-\epsilon_{33}\,.
\end{split}
\label{eq:constants}
\end{equation}
Here $\alpha_{\rm HH}$, $\gamma_{\rm HH}$, $\epsilon_{24}$ and $\epsilon_{33}$ are the box contributions from the double-hard expansions, while $r_2$ is the physical high-momentum tail at order $k^2$. The quantities $U$ and $W$ are the moments of $\tilde P(p)$ and $\tilde P(p)^2/p^2$ above $k_{\max}$, respectively. Like $S_{\rm UV}$ in eq.~\eqref{eq:seff}, they can be expressed as sums over the FFTLog coefficients $c_m$ and analytically continued. The rational coefficients arising from the overlap terms are
\begin{equation}
A_0=-\frac{481696}{3009825}\,,\qquad A_1=\frac{10541261}{463513050}\,,\qquad g=\frac{69056}{3565485}\,.
\label{eq:A01}
\end{equation}

The terms proportional to $\sigma^2_<$ describe the response to the low-momentum part of the spectrum. The one-loop response in eq.~\eqref{eq:ct1} is analytic, while the two-loop response $X(k)$ was evaluated numerically. For $\nu=-1.6$, both response terms are below $10^{-6}$ of the spectrum.

We now describe how the single-hard shapes, double-hard box contributions and overlap coefficients arise in the individual diagrams.

\emph{$P_{15}$.}
In the single-hard limit $p\gg k,q$, the angular average of $F_5(\kk,\qq,-\qq,\pp,-\pp)$ reduces to $G_3(\kk,\qq,-\qq)/p^2$. Since either loop momentum can be hard, this gives the factor of 60 in $S_{15}$. The large-$q$ expansion of the angular average of $60\,G_3$ is $k^2\big[A_0+A_1k^2/q^2+O(k^4/q^4)\big]$. These coefficients determine the overlap with the double-hard expansion at orders $k^2$ and $k^4$.

To evaluate the double-hard contribution, we consider $k\ll q\le p$ and expand the kernel in powers of the external momentum. Averaging $F_5$ over the direction of $\kk$ and over the angle between $\qq$ and $\pp$ gives
\begin{equation}
\frac{k^2}{p^2}\,\kappa(r)+\frac{k^4}{p^4}\,\kappa_4(r)+O(k^6)\,,\qquad r=\frac qp\,,
\label{eq:kappadef}
\end{equation}
with
\begin{equation}
\begin{split}
\kappa(r)={}&\frac{(1+r^2)(5760r^8+13605r^6-128258r^4+13605r^2+5760)}{103194000\,r^6}\\
&-\frac{(1-r^2)^4(384r^4+2699r^2+384)}{6879600\,r^7}\,\operatorname{arctanh}r\,.
\end{split}
\label{eq:kappa}
\end{equation}
The function $\kappa(r)$ is finite on $0\le r\le1$, with $\kappa(0)=A_0/60$ and $\kappa(1)=-11191/6449625$. In particular, its small-$r$ limit reproduces the single-hard coefficient, providing the matching needed for the overlap subtraction. The same function appears as the matter-density entry of the double-hard decomposition in eq.~(3.17) of ref.~\cite{Bakx:2026rmd}, where it is expressed as a combination of the functions $h_1$, $h_2$ and $h_3$.

The integral of this double-hard expansion over the FFTLog box gives
\begin{equation}
\alpha_{\rm HH}=-\frac{15}{2\pi^4}\sum_{m,n}c_mc_n\,\frac{k_{\max}^{a_m+a_n+4}}{a_m+a_n+4}\,\big[M(a_m)+M(a_n)\big]\,,\qquad
M(a)=\int_0^1\!\dd r\,r^{2+a}\,\kappa(r)\,,
\label{eq:alphaHH}
\end{equation}
where $M(a)$ is a Mellin moment of the double-hard kernel. It can be evaluated in closed form by writing $\kappa(r)=\sum_n\rho_nr^n+\operatorname{arctanh}(r)\sum_n\tau_nr^n$ and using
\begin{equation}
\int_0^1\!\dd r\,r^{s-1}=\frac1s\,,\qquad
\int_0^1\!\dd r\,r^{s-1}\operatorname{arctanh}r=\frac{\psi\big(\frac{s+1}{2}\big)-\psi\big(\frac12\big)}{2s}\,,\qquad s=a+n+3\,,
\label{eq:mellin}
\end{equation}
where $\psi$ is the digamma function. The poles of the individual terms cancel in the complete sum, while the poles of $M$ at $a=-3,-5,\ldots$ determine its analytic continuation.

The corresponding physical double-hard tail is $r_2$. It is obtained from the same double integral with $\plin(q)\plin(p)$ and $p>k_{\max}$, subtracting $\kappa(0)$ for $q<k_{\max}$ to remove the single-hard overlap. At order $k^4$, the same procedure applies to $\kappa_4$. Its small-$r$ behavior, $A_1/(60r^2)$, fixes the corresponding overlap and yields $\gamma_{\rm HH}$.

For $P_{24}$, the single-hard limit corresponds to the hard pair $\pp,-\pp$ entering $F_4$. Its angular average reduces to $G_2(\qq,\kk-\qq)/p^2$, giving the shape $S_{24}$. In the limit $q\gg k$, the angular average of $24F_2G_2$ behaves as $g\,k^4/q^2$, which determines the associated overlap contribution.

Unlike $P_{15}$, the double-hard expansion of $P_{24}$ starts at order $k^4$. It contains two powers of the spectrum evaluated at $q$ and one at $p$. Its integral over the FFTLog box gives $\epsilon_{24}$, which can be expressed as a triple sum over the coefficients $c_m$ of Mellin moments of the kernel $\kappa_{24}(q/p)$. The limiting behavior of this kernel is $g/r^2$ for $r\to0$ and $\bar\kappa_0/r^4$ for $r\to\infty$. These limits characterize the regions in which the hard momenta become strongly ordered.

The $P_{33}$ contributions have two distinct structures. In $P_{33\text{-}\rm II}$, the three loop momenta form a triangle. Consequently, there is no single-hard region and hence no overlap subtraction. Its leading double-hard contribution appears at order $k^4$, with a kernel that is a rational function symmetric in the three sides of the triangle. The corresponding box contribution, $\epsilon_{33}$, can be expressed as a triple sum of Beta functions.

The configurations of $P_{24}$ and $P_{33\text{-}\rm II}$ with exactly two hard loop momenta contain two powers of the spectrum at large momentum and are suppressed by $k_{\max}^{-2}$. These contributions are therefore subleading at the order considered here.

The remaining contribution, $P_{33\text{-}\rm I}$, is determined by the square of the one-loop correction. Using $P_{33\text{-}\rm I}=P_{13}^2/(4\plin)$ with $P_{13}=P^{\rm cut}_{13}-\tfrac{61}{315}S\,k^2\plin$, where $S=S_{\rm UV}$ or $S^{\rm phys}_{\rm UV}$ for the naive and physical calculations, respectively, gives the shape $S_{33}$ and the last term in $\gamma$ in eq.~\eqref{eq:constants}.

\begin{table}[t]
\centering
\begin{tabular}{lrrr}
\toprule
& $\nu=-1.6$ & $\nu=-1.8$ & $\nu=-0.8$\\
\midrule
$\alpha$ [$h^{-2}{\rm Mpc}^2$] & $2.0044$ & $1.9000$ & $-1887.8$\\
$\gamma$ [$h^{-4}{\rm Mpc}^4$] & $8\times10^{-6}$ & & $75.62$\\
$\epsilon$ [$h^{-7}{\rm Mpc}^7$] & $<10^{-8}$ & & $-343.0$\\
\bottomrule
\end{tabular}
\caption{The constants of eq.~\eqref{eq:constants} for the input spectrum of section~\ref{sec:fftlog} with $N=168$. At $\nu=-1.6$ the box term gives $94\%$ of $\alpha$, the overlap term $6\%$ and the physical tail a fraction $6\times10^{-5}$; the order-$k^4$ constants are suppressed by $k_{\max}^{-2}$ relative to $\alpha$, and $\epsilon$ by a further factor $\plin(k_{\max})$. At $\nu=-0.8$ the moments are continued and large, and all terms are needed.}
\label{tab:consts}
\end{table}

The quantities entering these expressions are available in closed form, except for $\kappa_4$, $\kappa_{24}$ and $\bar\kappa_0$, which were obtained from numerical angular averages with the exact asymptotic limits given above. The resulting coefficients are listed in table~\ref{tab:consts}.

For $\nu=-1.6$, the leading coefficient $\alpha$ is dominated by the double-hard box contribution, which accounts for $94\%$ of its value, while the overlap contributes $6\%$ and the physical high-momentum tail is negligible. The coefficients at order $k^4$ are much smaller, as expected from their additional suppression by powers of $k_{\max}$. For $\nu=-0.8$, the analytically continued high-momentum moments become large, and all contributions in eq.~\eqref{eq:constants} must be retained.

We have checked these coefficients against direct integration using the same input spectrum. For $\nu=-1.6$ and $-1.8$, subtracting $\alpha k^2\plin$ leaves a residual below $3\times10^{-8}\plin$ up to $0.2\hmpc$. For $\nu=-0.8$, all three coefficients agree with a fit to the residual.\footnote{On the interval $[10^{-4},50]\hmpc$ at $\nu=-0.8$, to $2\times10^{-9}$, $3\times10^{-5}$ and $5\times10^{-5}$ relative for $\alpha$, $\gamma$ and $\epsilon$. The neglected terms, higher orders of the single-hard expansion, the configurations of $P_{24}$ and $P_{33\text{-}\rm II}$ with two hard spectra at nearly opposite momenta, the physical double-hard tails at order $k^4$ and the effect of momenta below $k_{\min}$ on the overlap terms, are all below $10^{-7}$ in the units of table~\ref{tab:consts} or suppressed by $10^{-4}$ relative to the terms kept.}

\subsection{Counterterms at two loops}
\label{sec:ctlist}
Through two loops, $P_{\rm ct}$ contains three kinds of terms, eqs.~(4.82) and (4.83) of ref.~\cite{Anastasiou:2025jsy}: terms proportional to $k^2\plin(k)$, whose coefficient is related to the effective sound speed of the dark-matter fluid, and to $k^4\plin(k)$; one-loop integrals of the input spectrum in which one kernel $F_n$ is replaced by a counterterm kernel, with seven further coefficients; and a stochastic term, independent of $\plin(k)$, which scales as $k^4$ by mass and momentum conservation and which ref.~\cite{Anastasiou:2025jsy} neglects. Each coefficient absorbs the regularization-dependent contributions of large loop momenta to the loop integrals, and the terms of eqs.~\eqref{eq:ct1} and \eqref{eq:ct} have exactly these forms: the shapes are one-loop integrals with the large-momentum limit of a kernel in place of the kernel, with the moment as coefficient; $\alpha\,k^2\plin$ and $\gamma\,k^4\plin$ shift the sound-speed counterterm and its higher-derivative correction; and $\epsilon\,k^4$ has the form of the stochastic term.

\subsection{Comparison with ref.~\cite{Anastasiou:2025jsy}}
\label{sec:lit}
Ref.~\cite{Anastasiou:2025jsy} integrates over all loop momenta with the input spectrum of section~\ref{sec:fftlog}, computed to $10^4\hmpc$ and continued as $k^{-3}$, and subtracts from each integrand its limit for one or two large loop momenta above $M=0.6\hmpc$ (section~4 there). The subtracted terms have the forms of eqs.~\eqref{eq:ct1} and \eqref{eq:ct}, with the moment of the linear spectrum above $M$ in place of $\Delta S$,\footnote{As we read eq.~(4.37) of ref.~\cite{Anastasiou:2025jsy}, the term $S_{33}$ enters there with a factor two, while expanding $P_{33\text{-}\rm I}=P_{13}^2/(4\plin)$ to first order in the subtraction gives the factor one, which also follows from their subtracted $P_{33\text{-}\rm I}$ and their published values. We use the factor one.} and are absorbed by the counterterm coefficients; their published values are the subtracted corrections. The difference between our corrections and theirs is therefore again of these forms, with $\Delta S$ replaced by the moment of the power laws above $k_{\max}$ plus the moment of the linear spectrum between $M$ and $k_{\max}$. Our direct-integration code, run with their input and their subtractions, reproduces their published two-loop values within the combined Monte Carlo errors.

\begin{figure}[t]
\centering
\includegraphics[width=\textwidth]{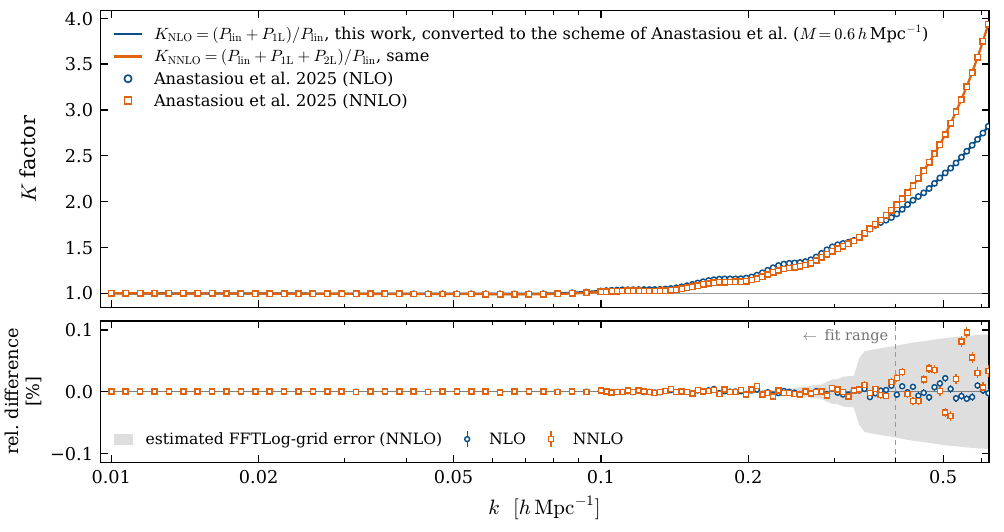}
\caption{The matter power spectrum through one loop (NLO) and two loops (NNLO), divided by the linear spectrum, at $z=0.57$ for the spectrum with BAO of ref.~\cite{Anastasiou:2025jsy}, without counterterms. Lines: this work, for $\nu=-1.6$, converted to the subtraction scheme of ref.~\cite{Anastasiou:2025jsy}; at one loop without free parameter, at two loops with the three constants $\alpha$, $\gamma$ and $\epsilon$ fitted for $k\le0.4\hmpc$ (dashed line). Symbols: the published values of ref.~\cite{Anastasiou:2025jsy}. Lower panel: relative difference; the band is the change of our result between 168 and 256 power laws (section~\ref{sec:validation}).}
\label{fig:bench}
\end{figure}

Here the external factor of $P_{13}$, $P_{15}$ and $P_{33\text{-}\rm I}$ is the input spectrum, and the displacement terms of appendix~\ref{sec:allchecks} are included. Figure~\ref{fig:bench} compares our result with their tables; neither calculation includes infrared resummation. At one loop the conversion has no free parameter and accounts for the difference up to $0.4\hmpc$; the remaining difference between $0.3$ and $0.6\hmpc$, up to $4\times10^{-4}$ of $P_{\rm 1L}$, is our reconstruction error. At two loops the three constants account for the difference up to $0.4\hmpc$, the fit range, and also describe the 16 points between $0.4$ and $0.6\hmpc$; leaving the coefficient of the shapes free as well returns its predicted value.\footnote{At one loop $\chi^2=72$ for 89 wavenumbers up to $0.4\hmpc$. The two-loop fit for $k\le0.4\hmpc$ gives $\alpha=36.3334(1)$, $\gamma=-0.141(9)$ and $\epsilon=-3(7)$ in units of $h^{-2}{\rm Mpc}^2$, $h^{-4}{\rm Mpc}^4$ and $h^{-7}{\rm Mpc}^7$, with $\chi^2=44.6$ for 86 degrees of freedom, below one per degree of freedom because our error band is conservative; with the coefficient of the shapes free, $2.732(21)$ against the predicted $2.717\,h^{-2}{\rm Mpc}^2$, and $\chi^2=44.1$ for 85. Before any conversion, our one-loop correction differs from theirs by up to $2.4$ times their value at low $k$, and our two-loop correction by up to $12\%$ of the linear spectrum between $0.3$ and $0.6\hmpc$.} After the conversion, the two results agree to $10^{-4}$ of the spectrum up to $0.3\hmpc$ and to $0.1\%$ up to $0.6\hmpc$, at the level of the grid error of section~\ref{sec:validation}. The comparison tests both calculations up to terms of the form $k^2\plin$, $k^4\plin$ and $k^4$, which the counterterms absorb.

\section{A single contribution: \texorpdfstring{$P_{33\text{-}\rm II}$}{P33-II}}
\label{sec:allchecks}

\begin{figure}[t]
\centering
\includegraphics[width=0.7\textwidth]{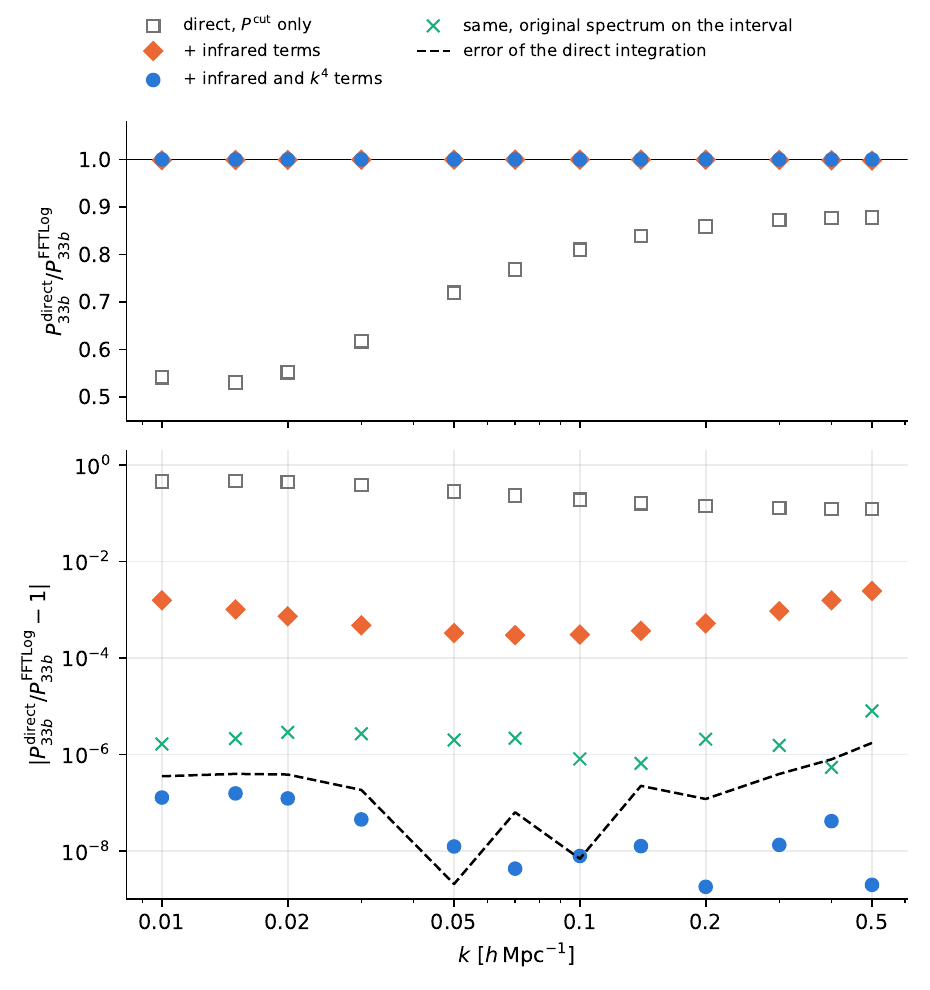}
\caption{$P_{33\text{-}\rm II}$ from direct integration, divided by the result from the precomputed FFTLog tensor, at 12 wavenumbers, for $\nu=-0.8$. Squares: direct integration with the spectrum on the FFTLog interval only, $P_{33\text{-}\rm II}^{\rm cut}$. Diamonds: with the infrared terms of eq.~\eqref{eq:p33bdiff} added. Circles: with all terms of eq.~\eqref{eq:p33bdiff} added; no free constant. Crosses (bottom only): the same as the circles, with the original linear spectrum in place of $P^{\rm cut}$ on the interval. Dashed line: error of the direct integration. Top: the ratio; bottom: its deviation from one.}
\label{fig:perdiag}
\end{figure}

This appendix gives, for a single contribution, a detailed example of how the loop momenta outside the FFTLog interval are treated. The comparison of section~\ref{sec:compare} needs the two-large-momenta terms of appendix~\ref{app:ctuv}; for a single contribution we can choose a bias at which they are absent. We take $P_{33\text{-}\rm II}$ of eq.~\eqref{eq:diagrams} at $\nu=-0.8$, where its integral converges: one soft loop momentum requires $\nu>-1$ and two large ones $\nu<-\tfrac23$, since $F_3$ grows as $1/q$ for a soft momentum and vanishes as $k^2/q^2$ for two large ones~\cite{Goroff:1986ep,Peebles:1980yev}. Then the analytic continuation drops nothing, and the naive result is the convergent integral of the whole sum of power laws, eq.~\eqref{eq:split}, including $P^<$ and $P^>$. We compare it with the direct integration of $P_{33\text{-}\rm II}$ with $P^{\rm cut}$ as input, $P^{\rm cut}_{33\text{-}\rm II}$. For $k_{\min}\ll k\ll k_{\max}$ the difference is
\begin{equation}
P^{\rm naive}_{33\text{-}\rm II}-P_{33\text{-}\rm II}^{\rm cut}=x\,P_{22}-\tfrac{1}{2}\,x^2\,\plin+\epsilon_{33}\,k^4\,,\qquad x=\frac{k^2S_{\rm IR}}{3}\,,
\label{eq:p33bdiff}
\end{equation}
up to corrections suppressed by powers of $k/k_{\max}$ and $k_{\min}/k$. The terms with $x$ come from the modes of $P^<$ below $k_{\min}$, which displace the matter uniformly~\cite{Jain:1995kx,Scoccimarro:1995if,Senatore:2014via,Baldauf:2015xfa}; $S_{\rm IR}$ is their displacement variance, the moment of $P^<$ with $1/p^2$ below $k_{\min}$. One soft line gives $x$ times the remaining one-loop integral, and two soft lines give the term with $x^2$. The term with $\epsilon_{33}$ comes from both loop momenta near $k_{\max}$, where the integrand expanded in $k$ gives $k^4$ times a function of the loop momenta, integrated with $P^{\rm cut}$; for the power laws the same region is scaleless and vanishes. A single large loop momentum gives no term, since every loop momentum of $P_{33\text{-}\rm II}$ enters two factors of the spectrum. Both $S_{\rm IR}$ and $\epsilon_{33}$ are finite sums over the coefficients $c_m$.

Figure~\ref{fig:perdiag} shows the comparison. The direct integration with $P^{\rm cut}$ alone gives between half and seven eighths of the FFTLog result; the infrared terms account for almost all of the difference, and the computed $\epsilon_{33}k^4$ for the remainder, up to $2\times10^{-3}$. With both, the two calculations agree at every wavenumber to better than $2\times10^{-7}$, at the level of the error of the direct integration. This tests the tensor, the FFTLog coefficients and both asymptotic terms together, without a free parameter. The crosses in figure~\ref{fig:perdiag} replace $P^{\rm cut}$ by the linear spectrum on the interval; the deviation, up to $10^{-5}$ at $0.5\hmpc$, is the effect of the reconstruction error on $P_{33\text{-}\rm II}$. The other contributions behave in the same way (section~\ref{sec:compare}).

\bibliographystyle{JHEP}
\bibliography{refs}

\end{document}